# On the Disturbance Equation of Flow Instability

Hua-Shu Dou

Temasek Laboratories,
National Wind Tunnel Building
National University of Singapore,
Singapore 117508,  SINGAPORE
Email: tsldh@nus.edu.sg; huashudou@yahoo.com

**Abstract**   In this note, we propose a new idea by analyzing the basic disturbance equations, and give starting equations for understanding the instability phenomena of laminar flows and transition to turbulence. It is considered that there is an interaction between the disturbance and the base flow in the process of disturbance amplification. The transition to turbulence is the result of this interaction. The linearity of the disturbance in linear theory is oversimplified the problem. It is pointed out that the development of disturbance is subjected to the influence of the base flow, and these effects must be included in the governing equations of the disturbance.

## 1. Introduction

Since Reynolds (1883) [1] famous experiments on pipe flow demonstrating the transition from laminar to turbulent flows, a large number of scientists have been investigating this for more than 120 years. Outstanding progress has been achieved in understanding the turbulent flow phenomena, predicting the properties of turbulence flows as well as developing the basic turbulent theory. However, the mechanism of turbulence generation still remains a great challenge today. Lumley and Yaglom (2000) [2] reviewed the advance of 100 years in turbulent researches, they concluded that it is very far from approaching a comprehensive theory and the final resolution of the turbulent problem. Various stability theories emerged during the past 100 years, but few are satisfactory in the explanation of the various flow instabilities and the related complex flow phenomena in the turbulence generation.

The origin of turbulence is a fundamental issue for the turbulence research. Most studies for the turbulence origin are to study the stability of a base flow under a disturbance. According to the stability theory, if the flow is stable, the disturbance will decay with the time; otherwise it will increase and the flow will be unstable. The unstable flow will evolve to a turbulent flow state eventually under the continuous disturbance.

Various stability theories have been developed during the past century. For example, linear stability analysis, energy method, week non-linear method, and second stability theory, etc. have been used in the researches [3-8]. Although some agreement is achieved between the theory and the experiments for some cases, the entire state-of-the-art in this topic is not satisfactory. In this note, we will propose a new approach to better understand the instability of laminar flow and transition to turbulence.

## 2. Linear Stability Theory

Linear stability theory exists from the year of Rayleigh (1880) [9] who did the inviscid stability analysis. This theory is still used now for the stability analysis, but the analysis includes the viscosity which solves the Orr-Sommerfeld equation. Linear stability analysis was successful for some flows such as Taylor-Couette flow, Dean flow, and Rayleigh-Benard problem, etc. But it is failed for wall bounded parallel flows such as Plane Poiseuille flow and pipe Poiseuille flow, and plane Couette flow [3-8]. The pipe Poiseuille flow (Hagen-Poiseuille) is stable by linear stability analysis for all the Reynolds number, Re. However, experiments showed that the flow would become turbulence if Re ($=\rho \bar{U} D/\mu$) exceeds a value of 2000 (see Table 1) [10]. Experiments also showed that disturbances in a laminar flow could be carefully avoided or considerably reduced, the onset of turbulence was delayed to Reynolds numbers up to Re=$O(10^5)$ [10]. For Re above 2000, the characteristic of turbulence transition depends on the disturbance amplitude and frequency, below which transition from laminar to turbulent state does not occur regardless of the initial disturbance amplitude. Linear stability analysis of plane parallel flows gives critical Reynolds number Re ($=\rho u_0 h/\mu$) of 5772 for plane Poiseuille flow [see 3-8], while experiments show that transition to turbulence occurs at Reynolds number of order 1000. For plane Couette flow, linear stability analysis shows that the flow is stable foe all the Re ($=\rho \bar{U} h/\mu$), but it becomes unstable for Re=370 in experiments [11-12]. One resolution of these paradoxes is that the domain of attraction for the laminar state shrinks for large Re (as $Re^\gamma$ say, with γ<0) in linear stability thoery, so that small but finite perturbations lead to transition [10-11]. However, this explanation does still not provide a good understanding for the problem and a satisfying prediction, compared to the experiments. Grossmann (2000) [12] commented that this discrepancy of linear stability analysis demonstrates that nature of the onset-of-turbulence mechanism in this flow must be different from an eigenvalue instability.

| Flow Type | Re expression | Linear Analysis | Experiments |
|---|---|---|---|
| Plane Poiseuille | Re= $\rho u_0 h/\mu$ | $Re_c$=5772. | $Re_c$ =1000. |
| Pipe Poiseuille | Re= $\rho \bar{U} D/\mu$ | $Re_c$ =∞. | $Re_c$ =2000. |
| Plane Couette | Re= $\rho \bar{U} h/\mu$ | $Re_c$ =∞. | $Re_c$ =370. |

Table 1 Critical Reynolds number for a few wall-bounded parallel flows. $\bar{U}$ is the averaged velocity, $u_0$ the velocity at the mid-plane of the channel, D the diameter of the pipe, h is the half-width of the channel for plane Poiseuille flow and plane Couette flow (Fig.1). From the comparison in this table, it is concluded that linear stability analysis is not liable for wall-bounded parallel flows.

Linear stability theory is based on the linearized equation of disturbance which is obtained from the Navier-Stokes equation. The presuppositions of this theory are that the disturbance equation for the resulting velocity **U**+**u** satisfies the Navier-Stokes equation, and the base flow **U** after overlapped by disturbance **u** still satisfies the Navier-Stokes equation. These hypothesizes are not proved yet so far and are of doubt although they are applied to many problems today and this theory represents the state-of-the-art of the stability theory.

### 3. Proposed New Equation

Let us consider that there is a steady laminar flow in a duct, which satisfy the Navier-Stokes equation. The conservations of mass and momentum for an incompressible Newtonian fluid (neglecting gravity force) can be expressed as,

$$\nabla \cdot \mathbf{U} = \mathbf{0} \tag{1}$$

$$\rho\left(\frac{\partial \mathbf{U}}{\partial t} + \mathbf{U} \cdot \nabla \mathbf{U}\right) = -\nabla P + \mu \nabla^2 \mathbf{U} \tag{2}$$

Now, we install a grid at the inlet of the duct to generate a small disturbance. We also assume that the flow rate is not reduced by the resistance of the grid. We think that the resultant flow field $\mathbf{V}=\mathbf{U}+\mathbf{u}$ may not be certain to satisfy the Navier-Stokes equation.

In order to carry out the analysis, now we assume that $\mathbf{U}+\mathbf{u}$ satisfies the Navier-Stokes equation,

$$\nabla \cdot (\mathbf{U} + \mathbf{u}) = \mathbf{0} \tag{3}$$

$$\rho\left(\frac{\partial (\mathbf{U} + \mathbf{u})}{\partial t} + (\mathbf{U} + \mathbf{u}) \cdot \nabla (\mathbf{U} + \mathbf{u})\right) = -\nabla (P + p) + \mu \nabla^2 (\mathbf{U} + \mathbf{u}). \tag{4}$$

Then, we decompose the governing equation for $\mathbf{U}$ and $\mathbf{u}$ into two equations,

$$\rho\left(\frac{\partial \mathbf{U_1}}{\partial t} + \mathbf{U_1} \cdot \nabla \mathbf{U_1}\right) = -\nabla P_1 + \mu \nabla^2 \mathbf{U_1} \tag{5}$$

$$\rho\left(\frac{\partial \mathbf{U_1}}{\partial t} + \mathbf{U_1} \cdot \nabla \mathbf{u_1} + \mathbf{u_1} \cdot \nabla \mathbf{U_1} + \mathbf{u_1} \cdot \nabla \mathbf{u_1}\right) = -\nabla p_1 + \mu \nabla^2 \mathbf{u_1}. \tag{6}$$

In these equations, the subscript 1 describes the difference of the meaning of the velocities which represented in Equation (5) and (6) from those in Equation (3) and (4). Here, assumption is made that $\mathbf{U_1}$ satisfied Navier-Stokes equation (i.e. Eq.(5)), the Eq.(6) is obtained by deducting the Eq.(5) from the Equation for $\mathbf{U}+\mathbf{u}$ (Eq.(4)). The linear stability analysis is usually carried out using Eq.(6) and dropping off the nonlinear term $\mathbf{u_1} \cdot \nabla \mathbf{u_1}$. Actually, we are still not sure that the base flow $\mathbf{U_1}$ exactly satisfies Navier-Stokes equation (5). In fact, the base flow and the disturbance flow may interact each other, and influence mutually. However, these interactions are ignored when Eq,(5) and (6) are constructed.

If we want to exactly analyze the flow using these two equations (Eq.(5) and (6)), necessary correction must be made. Now we rewrite Eq.(5) and (6) as by adding a correction function,

$$\rho\left(\frac{\partial \mathbf{U_1}}{\partial t} + \mathbf{U_1} \cdot \nabla \mathbf{U_1}\right) = -\nabla P_1 + \mu \nabla^2 \mathbf{U_1} + f_1(\mathbf{U_1}, \mathbf{u_1}) \tag{7}$$

$$\rho\left(\frac{\partial \mathbf{U_1}}{\partial t} + \mathbf{U_1}\cdot\nabla\mathbf{u_1} + \mathbf{u_1}\cdot\nabla\mathbf{U_1} + \mathbf{u_1}\cdot\nabla\mathbf{u_1}\right) = -\nabla p_1 + \mu\nabla^2\mathbf{u_1} + g_1(\mathbf{U_1},\mathbf{u_1}). \quad (8)$$

Here, $f_1$ and $g_1$ are two functions representing the effect from the opposition of $\mathbf{U}_1$ or $\mathbf{u}_1$, respectively. The function $f_1$ represents the influence of the disturbance $\mathbf{u}_1$ on the base flow $\mathbf{U}_1$, even it is small compared to the solution of Eq.(5) at small disturbance. The function $g_1$ is the effect of the base flow $\mathbf{U}_1$ on the disturbance flow, the influence of this term may not be small relative to the solution of Eq.(6). It may give an important Impact on the evolution of disturbance velocity $\mathbf{u}_1$.

The motivation to construct the Eq.(7) and Eq.(8) is based on the observation that the viscous force of the base flow, $\mu\nabla^2\mathbf{U_1}$, should have an important effect on the disturbance development [13-14]. It has a strong damping role to the disturbance. But, this influence disappears in Eq.(6) after dropping off the terms of the base flow. As is well known, there are two competing influences from the base flow with the variation of the Reynolds number, i.e., inertia and viscous forces. When the flow is subjected to a small disturbance, both inertia and viscous forces of the base flow have effects on this disturbance. However, in Eq.(6), only the partial effect of inertia of base flow which is from the nonlinearity remains and that of viscous force of base flow vanishes. This is obviously not reasonable. The stability of a laminar flow might be much different from the mechanical analogue of a simple linear oscillator. The evolution of disturbance in laminar flow is encompassed and affected by the media of viscous flows. But, the motion of a linear oscillator is free of the influence of the environmental media under the given existing forces. The treatment of linearity for disturbance in linear theory is oversimplified the problem. To compensate the simplification of the equation, it should not be to simply account for the non-linearity due to the inertia, both the effects of the viscous force and the inertia force of the base flow should be included.

Only we solve Eq.(7) and (8), can we get the solution which satisfies the physical phenomena exactly. However, the determination of $f_1$ and $g_1$ needs further extensive researches, and it seems not easy. This note provides a starting point for understanding this complicated physical phenomenon.

## 4. Concluding Remarks

The development of disturbance in fluid flows is different from those of solid mechanics. The stability of a laminar flow might be much different from the mechanical analogue of a simple linear oscillator. In fluid flow, there is interaction between the base flow and the disturbance flow. The process of linearity of disturbance is over-simplified the problem. The solution of linear eigenvalue is correct in mathematics for the given equation, but it is not true in physics; even although linear analysis could give agreement result with experiment in some cases (for example, Tollmien-Schlichting waves in boundary layer and Taylor-Couette flow between rotating cylinders). This is because the starting equations do not represent the physics of the problem. In the solution of the disturbance equation, only the interaction of the base flow and the disturbance flow is considered, results consistent with the physics could be obtained.

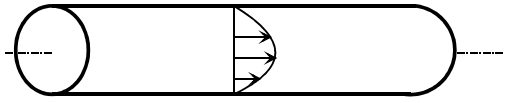

(a)

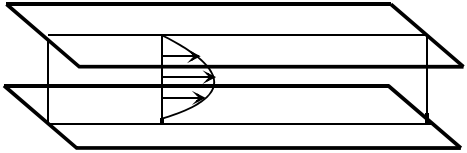

(b)

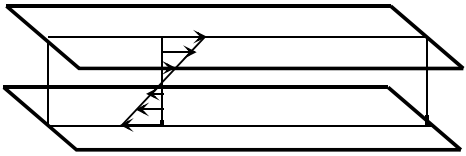

(c)

Fig.1 Parallel flows. (a) Pipe Poiseuille flow; (b) Plane Poiseuille flow; (c) Plane Couette flow.